# Elasticity in the Gauge Theory of Active Nematics with Topological Defects

*L.V. Elnikova[1*]*

[1]*A.I. Alikhanov Institute for Theoretical and Experimental Physics - NRC "Kurchatov Institute", Bolshaya Cheremushkinskaya 25, Moscow 117218, Russian Federation*

[*]*e-mail: elnikova@itep.ru*

We analyze the phase behavior of lyotropic nematic liquid crystals in the self-organizing flow called usually by active nematics (AN). Their elastic properties are mutually caused by evolution of topological defects (for instance, disclinations and boojums) and the flow regime. Such changes in elasticity of AN comparing with conventional inactive ones set the new working characteristics of these materials. These changes have an influence on switchable and tunable properties of AN. In this work, we study the phases of a droplet composed of uniaxial AN with topological defects in their collective flow. We apply the gauge string-like theory using the method of differential forms on a lattice interchanging the drive-force concept. The results of our numerical modeling with Monte Carlo method show, that under certain conditions, the type of the phase transition from nematic to isotropic (N-I) phase and the thermodynamical characteristics in an active regime may differ from such one in the conventional lyotropic nematics.

1. Introduction

Active nematics are classified as multicomponent (lyotropic) nematic liquid crystalline (NLC) systems, exhibiting collective behavior and spontaneous flow [1-5]. These are novel metamaterials, Janus particles, active colloids, Quincke rotators in isotropic, nematic or smectic-A hosts, granular biological particles, etc. [4-7]. Because of distortion or reorientation of the order parameter under external factors (electric or magnetic fields *etc*.), active nematics possess switchable and tunable physical characteristics and are used as perspective materials in modern photovoltaic devices, sensors and actuators, *etc*.

In the theory of dielectric relaxation for nematic liquid crystals (NLC), there is a representation of director dynamics controlled by the balance of dielectric, viscous, and elastic torques through the Ericksen-Leslie equation [8], going back to the pioneering works by Saupe [9-11]. In the terms of the Green, Toner, Vitelli (GTV) and similar approaches [2-5], individual "birds" may be associated with local spin variables of the classical *XY* model [3] in long-range ordering, and the dynamics nonpotential.

On the other hand, there is the dynamical gauge theory of linear defects proposed by Kadić and Edelen (KE) [12], in this representation, the defect structure as a gauge field evolves via Lagrangians of the non-abelian group *SO*(3) with semidirect product as well as the translation group. The colloid phases of active nematic liquid crystals (ANLC) possess such topological defects, so that the GTV model has to be specified with a glance of symmetry and a non-trivial action of its group.

In this work and similar to the Yang-Mills approach [13], we develop the representation of configurations with defect like boojums and disclinations condensate in ANs, expressed on terms of monopoles and disclinations of the gauge theory. We make an attempt to spread out these concepts onto the Frank energy [14] with the active force in hydrodynamics in the sense GTV [2]. We provide the numerical modeling for the observables, thermodynamical values and correlation functions, using the formalism of differential forms on a lattice.

## 2. Dynamics and The Gauge Theory

Due to the continual Ericksen-Leslie theory [8,14] the hydrodynamical model of a flowing nematic includes the Navier-Stokes (NS) and nematodynamic equations, which are respectively [2,4]:

$$\rho_0 \frac{\partial v_k}{\partial t} = -\partial_k P + \eta \nabla^2 v_k + \alpha \partial_j (n_j n_k) + \partial_j \left( \lambda_{ijk} \frac{\delta F}{\delta n_i} \right), \tag{1.1}$$

$$\nabla \bullet \mathbf{v} = 0, \tag{1.2}$$

and 
$$\frac{\partial n_i}{\partial t} = \lambda_{ijk} \partial_j v_k - \frac{1}{\gamma_1} \left[ \frac{\delta F}{\delta n_i} - \left( \frac{\delta F}{\delta \hat{n}} \bullet \hat{n} \right) n_i \right], \tag{1.3}$$

where the material time derivatives of the director and velocity vector fields are given by

$$\frac{Dv_k}{Dt} = \frac{\partial v_k}{\partial t} + v \bullet \nabla v_k, \quad \frac{Dv_n}{Dt} = \frac{\partial n_k}{\partial t} + v \bullet \nabla \mathbf{n} \tag{1.4}$$

and the tensor $\lambda_{ijk}$ is defined by

$$\lambda_{ijk} = \left( \frac{\lambda+1}{2} \right) n_j \delta_{ik} + \left( \frac{\lambda-1}{2} \right) n_k \delta_{ij} - \lambda n_i n_j n_k . \tag{1.5}$$

The usual summation convention is adopted and $v(r,t)$ is the velocity field (the typical speeds of active particles are approximately (10-40) μm/s), $r$(x,y,z) is the position vector which depends on Cartesian coordinates x, y, and z, t is the time and $n(r,t)$ is a director vector field. The equation (1.2) is the incompressibility condition which is deduced from the continuity equation assuming that the mass density is $\rho_0$ is the constant. $P$ is the dynamical pressure, $\frac{\delta F}{\delta \hat{n}}$ is the molecular field, $\delta_{ij}$ is the Kronecker symbol, $\eta$ is the shear viscosity which is assumed to be isotropic for simplicity and $\gamma_1$ is the director field rotational viscosity given by $\gamma_1=\alpha_3-\alpha_2$ with $\alpha_2$ and $\alpha_3$ the Leslie viscosity coefficients. The dimensionless flow-alignment parameter $\lambda$ in equation (1.5) captured the anisotropic response of the nematogens to shear. The third term on the right hand side of equation (1.1) is the active force component in the $k$-axis direction ($k =1, 2, 3$) which may contractile if the activity parameter $\alpha > 0$ or extensile if $\alpha < 0$ depending on the system. The active force can be written as $\mathbf{F}_a = \alpha[\mathbf{n}(\nabla \bullet \mathbf{n}) + (\mathbf{n} \bullet \nabla)\mathbf{n}]$ which is produced by director distortions.

The Oseen-Frank elastic energy $F$ of a uniaxial nematic is given by

$$F = \frac{1}{2}\int_V \left[ K_1(\nabla \bullet \mathbf{n})^2 + K_2(\mathbf{n} \bullet (\nabla \times \mathbf{n}))^2 + K_3|\mathbf{n} \times (\nabla \times \mathbf{n})|^2 \right] dV \qquad (2)$$

which includes the elastic constants $K_1$, $K_2$ and $K_3$ associated with the splay ($\nabla \bullet \mathbf{n}$), twist ($\mathbf{n} \bullet (\nabla \times \mathbf{n})$), and bend ($\mathbf{n} \times (\nabla \times \mathbf{n})$) terms in (2), respectively. The volume of the region occupied by the active nematic is denoted by $V$. Here, we adopt the one elastic constant approximation, i.e., $K_1 = K_2 = K_3 = K$ for simplicity [14].

The model described by equations (1.1), (1.2), (1.3), (1.4), (1.5) and (2) from the point of view to a molecular field, is based on the unique combination of the material parameters expressing viscous, inertial and elastic effects in terms of the Ericksen or Reynolds numbers and the Leslie coefficients [15,16,17].

The notion of dynamical disclinations in the gauge approach (Kadić and Edelen, 1983 [12]) allows us to study also the elastic properties of a defective material, which are nonlinear in origin. The goal of the gauge theory is to transform the concept of force by the geometrical view of the bundle connectivity. This is naturally enough if we have in mind the presence of flow induced by evolution of topological defect configurations in individual self-organizing nematic colloids. In defect dynamics and due to the KE theory, the deformation gradient matrix must be replaced by the distortion tensor. Based on the KE approach, the monopole-like analytical solution for the disclination associated with the *SO*(3) Lagrangian was presented by Osipov [18].

The non-abelian rotation group *SO*(3) of the nematic director Lagrangian is locally isomorphic to *SU*(2). In the 3D abelian projection to *U*(1) [19,20], we have the Lagrangian

$$L = \frac{1}{g^2} G_{\mu\nu}^2 + |D_\mu \Phi|^2 + \lambda_2 \left(|\Phi|^2 - 1\right)^2, \tag{3}$$

with the monopole field $\Phi$, $D_\mu = \partial_\mu + iB_\mu$. $G_{\mu\nu} = \partial_\mu B_\nu - \partial_\nu B_\mu$, $B_\mu$ is the dual gauge field analogue to the electromagnetic field which is given by $G_{\mu\nu} = \frac{1}{2} \varepsilon_{\mu\nu\alpha\beta} (\partial_\alpha A_\beta - \partial_\beta A_\alpha)$ with $\varepsilon_{\mu\nu\alpha\beta}$ the Levi-Civita symbol and $A_i$ is the ordinary gauge field. The field $G_{\mu\nu}$ is addressed to the dissipative and whole kinetic energy. Note that this field as well as the Ericksen number may be introduced from (1) and (2) as the parameters. The monopoles interact by a flux tube via the Coulomb potential identically to a disclination with a core. The partition function for monopole currents between two boojums in an individual nematic droplet can be formulated in terms of differential operators.

3. **Method of differential forms, Monte Carlo results and discussions**

To characterize the evolution of a topological defect in an AN, we study the properties of the partition function for currents *j* in the form ([19] and references therein)

$$Z = const \sum_{\substack{*j \in Z(C_{k+2}) \\ \delta j = 0}} \exp\left\{-4\pi^2 \beta (j, \Delta^{-1} j)\right\}. \tag{4}$$

Here the sum runs over the (*D*-*k*-2)-forms of monopole currents *j* as the sum over the monopole defects (here, boojums) of the original theory, here *k* is the rank of differential form, $C_k$ is a *k*-dimensional cell of the lattice, $\Delta$ is the Laplace operator on the operators of a dual lattice. In this notation, at *k*=0, *D*=3, we have the *XY* model with the standard action $1/T\sum_{x,\mu} \cos(\varphi_x - \varphi_{x+k})$, where *x* are the lattice sites, $\varphi_i$ are compact dynamical variables running from $-\pi$ to $\pi$, $i=1\ldots 4$, when we follow the plaquettes of the dual cubic lattice. The dual currents calculated in modulo $2\pi$ are $*j = \frac{1}{2}\pi (|\varphi_1-\varphi_2|_{2\pi}+ |\varphi_2-\varphi_3|_{2\pi}+ |\varphi_3-\varphi_4|_{2\pi}+ |\varphi_4-\varphi_1|_{2\pi})$.

The Figure 1 shows one of the observable thermodynamic values, the specific heat, which is calculated at the dual lattice with self-dual conditions as well as at the gauge field $A_i$, reflected in the hydrodynamic parameter the Ericksen number $Er = \gamma_1 v r_0/K$ where $\gamma_1$ is the director field rotational viscosity which is given by $\gamma_1 = \alpha_3 - \alpha_2$ with $\alpha_3$ and $\alpha_2$ the Leslie coefficients. The value $v$ is the velocity in the position $r_o$ in the relations (1.1-1.5). The second standard parameter of the model is the chemical potential.

The expected transition from nematic to isotropic phase (N→I) corresponds to the transition 'confinement-deconfinement' in the string theory terms. In comparison with the conventional nematics [21], this transformation (Fig.1) is shifted to high temperatures showing a smooth extremum and some threshold character at increasing temperature.

## 4. Conclusions

In this work, we demonstrated strong validity of the gauge theory for the dissipative systems with the non-abelian order parameter at the field of topological defects, i.e., for the case of AN liquid crystals, where the drag force is unified in the gauge field.

In consequence of applications of this gauge and differential form methods, we can control in principle the hydrodynamical values in absent of some measured data in the flow regime, which is important for predictions of the phase behavior and elastic properties of novel artificial materials.

We demonstrated that the typical N→I phase transition in colloid AN droplets differ from that one in inactive conventional lyotropic nematics.

## 5. References

1. Vicsek T, Czirok A, Ben-Jacob E, Cohen I, and Shochet O. Novel Type of Phase Transition in a System of Self-Driven Particles. *Phys. Rev. Lett.* 1995;75:1226.

2. Green R, Toner J, Vitelli V. The geometry of thresholdless active flow in nematic microfluidics. ArXiv:1602.00561v3.

3. Toner J, Tu Y. Long-Range Order in a Two-Dimensional Dynamical XY Model: How Birds Fly Together. *Phys. Rev. Lett.* 1995;75:4326-4329.

4. Ryskin G, Kremenetsky M. Drag Force on a Line Defect Moving through an Otherwise Undisturbed Field: Disclination Line in a Nematic Liquid Crystal. *Phys. Rev. Lett.* 1991;67:1574–1577.

5. Marchetti MC, Joanny JF, Ramaswamy S, Liverpoo TB, Prost J, Rao M, Aditi Simha R. Hydrodynamics of soft active matter. *Rev. Mod. Phys*. 2013;85:1143-1189.

6. Lavrentovich OD. Active colloids in liquid crystals. *Curr. Opin. Colloid Interface Sci*. 2016;21:97-110.

7. Zhou S, Sokolov A, Lavrentovich OD, Aranson IS. Living liquid crystals. *Proceedings of the National Academy of Sciences*. 2014;111:1265.

8. Leslie FM. Some constitutive equations for anisotropic fluids, *Quart. J. Mech. Appl. Math.* 1966;19:357-370.

9. Maier W, Saupe A. One of the first applications of the molecular field theory developed by Maier and Saupe. *Z. Naturforsch*. 1958;13a:564-566.

10. Maier W, Saupe A. A simple molecular-statistics theory of the nematic liquid-crystalline phase, part I. *Z. Naturforsch.* 1959;14a:882-900.

11. Maier W, Saupe A. A simple molecular-statistics theory of the nematic liquid-crystalline phase, part II. *Z. Naturforsch*. 1960;15a:287-292.

12. Kadić A, Edelen DGB. *A Gauge Theory of Dislocations and Disclinations*. Berlin Heidelberg: Springer; 1983.

13. Yang CN, Mills R. Conservation of Isotopic Spin and Isotopic Gauge Invariance. *Phys. Rev.* 1954; 96:191-195.

14. de Gennes P, *The Physics of Liquid Crystals*, Oxford: Clarendon Press; 1974.

15. Landau LD, Lifshitz EM. *Gidrodinamika* V. 6, Moscow: Nauka; 1988.

16. Sengupta A. *Topological Microfluidics. Nematic Liquid Crystals and Nematic Colloids in Microfluidic Environment*; Springer: International Publishing Switzerland, 2013.

17. Zakharov AV, Vakulenko AA. Dynamics of the modulated distortions in confined nematic liquid crystals. *The J. of Chem. Phys.* 2013;139:244904-1–244904-6.

18. Osipov VA. A Monopole-like solution for static disclinations in continuum media. *Phys. Lett. A* 1990;46:67-70.

19. Polikarpov MI. Fractals, topological defects and confinement in lattice gauge theory. *Physics-Uspekhi*. 1995;165:627–644.

20. Ivanenko TL, Polikarpov MI, Pochinsky AV. Condensate of monopoles and confinement in an SU(2) lattice gauge theories. *JETP Lett*. 1991; 53: 543-545.

21. Skačej G, Zannoni C. Biaxial liquid-crystal elastomers: A lattice model. *Eur. Phys. J. E*, 2008;25:181-186.

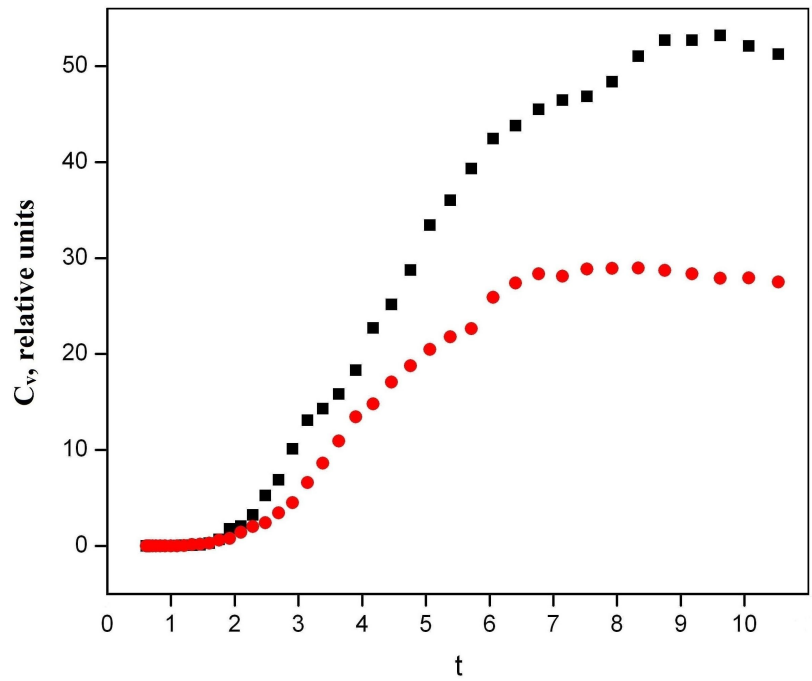

**Figure captures**

Fig. 1. Temperature dependence of specific heat for two simple cubic lattice sizes of $48^3$ (black solid squares) and $64^3$ (red solid circles), calculating errors are not shown. The *Er* number takes the value of 10 which is chosen arbitrary and the chemical potential is equal to 0.1.